\documentclass
[a4paper,twocolumn,showkeys,superscriptaddress,prb]
{revtex4}
\usepackage{graphicx}
\usepackage{natbib}
\usepackage{epsfig}
\usepackage{float}
\usepackage{amsmath, amsfonts, amssymb}
\usepackage{palatino}
\usepackage{flafter}
\usepackage{psboxit}
\usepackage{psfrag}
\usepackage{ulem}
\usepackage[all,cmtip]{xy}

\makeatletter
\def\@dotsep{4.5}
\makeatother

\bibstyle{jcp}
\newcommand{\expt}[1]{\langle #1 \rangle}

\begin{document}
\normalem

\title{
    Coarse-Graining and Scaling in Dissipative Particle
    Dynamics
}
\author{Rudolf M. F\"uchslin}
\affiliation{Ruhr Universit\"at Bochum, Biomolecular
    Information Processing (BioMIP), c/o BMZ,
    Otto-Hahnstr. 15, D-44227 Dortmund, Germany
}
\author{Harold Fellermann}
\affiliation{
	Self-Organizing Systems, EES-6 MS-D462, Los Alamos National Laboratory,
    Los Alamos 87545 NM, USA
}
\affiliation{
    ICREA-Complex Systems Lab, Universitat Pompeu Fabra
    (GRIB), Dr Aiguader 88, 08003 Barcelona, Spain
}
\author{Anders Eriksson}
\affiliation{
	Department of Energy and Environment,
    Chalmers University of Technology,
    SE-41296 G\"oteborg, Sweden
}
\author{Hans-Joachim Ziock}
\affiliation{
	Self-Organizing Systems, EES-6 MS-D462, Los Alamos National Laboratory,
    Los Alamos 87545 NM, USA
}
\thanks{
    Authors for correspondence: rudolf.fuechslin@biomip.ruhr-uni-bochum.de
}

\begin{abstract}
Dissipative particle dynamics (DPD) is now a well-established method for
simulating soft matter systems. However, its applicability was recently
questioned because some investigations showed an upper coarse-graining
limit that would prevent the applicability of the method to the whole 
mesoscopic range. This article aims to reestablish DPD as a truly
scale-free method by analyzing the problems reported by other authors and 
by presenting a scaling scheme that allows one to gauge a DPD-simulation 
to any desired length scale.
\end{abstract}

\maketitle

\section{Introduction}
Dissipative particle dynamics (DPD) was introduced in 1992 by Hoogerbrugge
and Koelman\cite{Hoo:1992} as a novel method for performing mesoscopic
simulations of complex fluids. Since then, the method has gained 
significant theoretical support and
refinement,\cite{Esp:1995, Gro:1997, Tro:2002, Esp:2005} and has
been applied to fluid dynamics in numerous research areas such as
rheology,\cite{Hoo:1992, Koe:1993, Sim:2004, Mar:2005} material
sciences,\cite{Kon:1997, Kon:1997a} and molecular biology, where
membranes,\cite{Ven:1999, Gro:2001} vesicles,\cite{Yam:2002, Yam:2003} and
micellar systems \cite{Gro:2000, Gro:2003, Fel:2006} have been modeled.

Initially, DPD was understood to be a truly mesoscopic method able to bridge
the whole gap between the underlying atomistic scale (in the range of
nanometers and nanoseconds) that is accessible by molecular dynamics (MD)
simulations and the macroscopic scale (in the range of micrometers and
milliseconds) considered by continuum descriptions. To fulfill this promise,
it is crucial that the method is scalable, meaning that its coarse-graining
level can be adjusted without introducing serious artifacts that would
render the method worthless.

Coarse-graining in DPD translates into having a number $\nu$ of physical
molecules be represented by a single DPD particle.\cite{Gro:2001} 
By $N$, we denote the  total number of DPD particles in a simulation and 
it holds $\nu N = N_\text{phys}$, with $N_\text{phys}$ the number of 
physical molecules the simulation represents.
The main objective of this article is the 
comparison of DPD simulations with different coarse-graining levels 
$\nu$ and $\nu'$. This motivates the introduction of the scaling ratio
$\phi = N/N' = \nu'/\nu$. In the following,
functions of $\phi$ will be used to describe the scaling of various 
quantities at different coarse-graining levels.
(One may set $\nu=1$ so that scaling ratio and coarse-graining level
coincide and thereby simplify the notation and providing a direct link to
physical constants such as the Boltzmann constant; this is at the price of
using DPD concepts at a length scale on which their interpretation
becomes difficult.)

It was originally stated that the method is scale-free, meaning
that the parameters used in the simulation do not depend on
the level of coarse-graining.\cite{Gro:1997} In a later 
publication,\cite{Gro:2001} this earlier finding was declared erroneous,
and it was proposed that interaction parameters determining the
conservative forces between DPD particles scale linearly with $\phi$
when one goes from one coarse-graining level $\nu$ to another $\nu'$.

Based on this scaling relation, the performance of DPD was
analyzed for various coarse-graining levels.\cite{Dzw:2000,Tro:2003}
It was found that there exists an upper coarse-graining level
above which the simulated fluid freezes. Trovimof reported that this
coarse-graining limit is disappointingly low,\cite{Tro:2003} and
only allows up to about 10 water molecules to be grouped together
into one DPD particle. This limit would prevent DPD from covering the 
whole mesoscopic range and confines its applicability essentially to the
order of magnitude of MD simulations. Dzwinel and Yuen even
concluded that the DPD method would be best suited for the simulation
of vapors and gases (where the freezing artifact would happen only
for much higher coarsening levels).\cite{Dzw:2000}

In this article, we argue that the original statement that DPD is scale 
free is in fact correct, and that the later refinement declaring that the 
DPD interaction parameter scales linearly with the scaling ratio $\phi$ is
based on an inappropriate scaling assumption. More specifically, we show
that if a scaling method that preserves physical quantities is used and the
DPD calculations are performed in reduced units (which is standard practice)
then the results are independent of the level of coarse-graining.
Expressed in more technical terms, if a physically consistent
scaling of all simulation parameters is maintained, the velocity increments
calculated from integrating the equations of motion of the DPD particles
(c.f. Eqns.~(\ref{equ_basicDPD}) below) do not depend on the
coarse-graining value used for the DPD particles. This is in contrast to 
earlier publications.\cite{Gro:2001,Gro:2004} 

In this paper, the physical molecules will refer to builk fluid particles
(such as water), which is in accordance with the cited literature. 
Systems including surfaces, such as binary fluids, will be commented on 
in the discussion.

\begin{figure}
    \centering
    \includegraphics[width=\columnwidth]
        {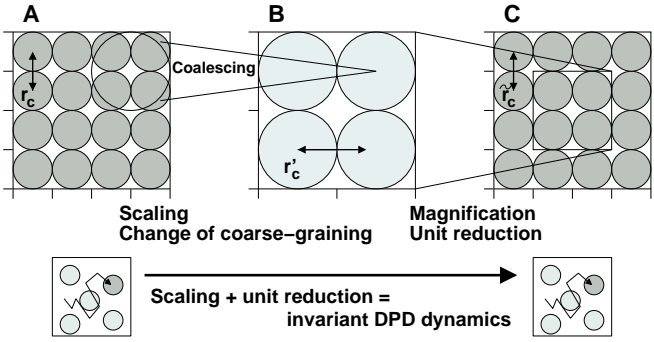}
    \caption{
Starting from Frame~A, a renormalization or scaling is performed
that changes the interaction parameters, but does not change the units,
leading to Frame~B. This step also involves changing some of the force
parameters as well a system constant (namely the Boltzmann factor) in
order to maintain the same behavior as the initial system had.
The overall system scaling is then changed with parameters being
expressed in terms of reduced units. The dynamics in the reduced unit
system (Frame~C) is numerically exactly the same as in Frame~A. In fact,
Frame~C is effectively a zoomed version of Frame~A.
}
\label{fig_renorm_and_blowup}
\end{figure}

The article is structured as follows: in Sec.~\ref{sec_GW}, we discuss
the results of Groot and Rabone\cite{Gro:2001} and specify where we
deviate from their analysis. We show that their approach of decreasing
the number of DPD particles (namely the particle density) while 
keeping relevant properties, in particular their radius of 
interaction, constant, is not appropriate.
The alternative scaling process we employ is schematically shown in
Fig.~1A and 1B. When we change the level of coarse-graining for
the DPD-particles, we accordingly scale their number and 
adjusting their size (radius of interaction). Fig.~1C depicts the 
main result to be shown in this article, namely that by employing the 
correct scaling relations and unit reduction one has a complete 
equivalence of a simulation performed at the scale of Fig.~1A with the 
zoomed version in Fig.~1C. In other words, we demonstrate that one 
single DPD-simulation represents the dynamics of a whole family of 
physical systems at different levels of coarse-graining. 

This demonstration is split into two parts. In
Sec.~\ref{sec_renorm_compress} and \ref{sec_renorm_heat}, we change the 
interaction parameters according to the necessities of the scaling 
procedure we adopt. The interaction parameters have do be changed 
such that a system with many DPD-particles is mapped onto one
with fewer, but larger and heavier particles. This is done without
changing the physical properties of the complete system, namely the
compressibility and thermal energy. Sec.~\ref{sec_simulation} presents
simulation results that corroborate the correctness of the
derived scaling relations.

In Sec.~\ref{sec_reduced_units}, we analyze the behavior of the
DPD algorithm when the rescaled system is expressed in its own set of 
reduced units. The change of units 
affects the time scale and we will show that as a consequence, the 
numerical values that appear in the reduced unit system 
(Fig.~\ref{fig_renorm_and_blowup}C) are identical to those of the original 
one and, as a consequence, their dynamics are equivalent.

In order to present our argument, we need to define the nomenclature used
and introduce some notation. By ``coarse-graining'' we understand the 
operation of coalescing $\nu$ physical particles into one DPD particle. 
By ``scaling'' we refer to the functional relation between the respective 
parameters of two systems with different coarse-graining resolutions.

Conventionally, DPD operates in reduced units, such that energy is measured
in units of $kT$ ($T$ being the temperature and $k$ a scaled version of 
the Boltzmann constant to be introduced later), length in units of $r_c$, 
and mass in units of $m$, the mass of a single particle; in these units, 
length, mass, time and energy are dimensionless. We will use two different 
sets of reduced units, one for the coarse-graining resolution $\nu$, and 
another one for $\nu'$. We denote by $X$ a quantity 
expressed in reduced units with respect to coarse-graining level $\nu$. 
With $X'$ we denote the corresponding value of the same 
quantity calculated 
in the units of $X$ but with a different level of coarse-graining.
In other words: moving from $X$ to $X'$ denotes the scaling operation.
Finally, by $\tilde X$ we refer to the quantity $X'$ in its own set of 
reduced units given by the coarse-graining resolution $\nu'$.
That is, going from $X'$ to $\tilde X$ denotes the reduction of units.
One may summarize the combined coarse-graining and change of units with 
the following diagram:
\[
\large
\xymatrixcolsep{6pc}
\xymatrixrowsep{3pc}
\xymatrix{
    X 
        \ar[r]^{\substack{\text{Scaling}\\ \text{by factor $\phi$}}} 
        \ar@{.}[rd]_{=}
    & X' 
        \ar[d]^{\substack{\text{Reduction}\\ \text{of units}}} \\ 
    & \tilde X
}
\]

Excellent descriptions of the DPD-method are given in various articles;
we will not recapitulate the method itself, but instead refer
to Groot and Warren.\cite{Gro:1997}
Here, we give only the definitions of  the conservative, dissipative, and 
random forces in order to define the notation of the parameters:
\begin{align}\label{equ_basicDPD}
	\mathbf{F}_{ij}^C
	    &= a_{ij} \chi_{ij} (1 - \frac{r_{ij}}{r_c})\mathbf{\hat{r}}_{ij}, 
	    \nonumber\\
	\mathbf{F}_{ij}^D 
	    &= - \gamma \omega^D(r_{ij}) 
	    \left[ (\mathbf v_i - \mathbf v_j) \cdot \mathbf{\hat{r}}_{ij} 
	    \right] \mathbf{\hat{r}}_{ij}, \nonumber\\
	\mathbf{F}_{ij}^R 
	    &= \sigma \omega^R(r_{ij}) \zeta_{ij} \mathbf{\hat{r}}_{ij},
\end{align}
where $r_{ij}$ is the Euclidean distance between particles $i$ and $j$,
$\mathbf{\hat{r}}_{ij}$ is the unit vector pointing from particle $j$ to
particle $i$, and  $\chi_{ij}$ equals one for pairs of particles separated 
by distances less than the force cut-off radius $r_c$ and equals zero 
otherwise. The
parameter $a_{ij}$ determines the magnitude of the conservative interaction
and will be regarded in this work as being the same for all pairs of
particles: $a_{ij} = a$.
$\zeta_{ij}$ is a random variable with Gaussian statistics, 
a vanishing mean and a variance of $1/\Delta t$ for the numerical time step
$\Delta t$
(see e.g.~Gardiner\cite{Gar:2004}). As a consequence, the unit of
$\zeta_{ij}$ is $\text{time}^{-1/2}$. Furthermore, the values of
$\zeta_{ij}$ in two different time intervals are uncorrelated. In general, 
if $(i,j)$ and $(k,l)$ are different pairs of particles, $\zeta_{ij}$ and
$\zeta_{kl}$ are independent; however, in order to guarantee the centrality
of all forces, one must require that $\zeta_{ij}=\zeta_{ji}$.

The following dissipation-fluctuation relation\cite{Esp:1995} leads to
a thermal equilibrium at a given temperature $T$:
\begin{equation}
    \label{eq_dissipation_fluctuation}
	2 k_B T \gamma \omega^D(r) = \sigma^2 [\omega^R(r)]^2.
\end{equation}
Without loss of generality, we may take $\omega^D(r) = [\omega^R(r)]^2$ 
for the dimensionless weighing functions $\omega^D(r)$ and $\omega^R(r)$. 
The dissipation-fluctuation relation~(\ref{eq_dissipation_fluctuation}) then
reduces to
\begin{equation}
    \label{eq_dissipation_fluctuation2}
	\sigma^2 = 2 k_B T \gamma.
\end{equation}
One is free to choose either $\omega^D(r)$ or $\omega^R(r)$ without 
changing the thermodynamic equilibrium, but it is customary in the
literature to take $\omega^R(r) = 1 - r/r_c$ mimicking the 
conservative force $\mathbf F^C$. One also notes that through 
Eqns.~(\ref{eq_dissipation_fluctuation}) and 
(\ref{eq_dissipation_fluctuation2}), the relations between the
parameters therein are dependent on the temperature.

\section{Scaling DPD}
\label{sec_StatmechGW}

\subsection{Compressibility and Equation of State}
\label{sec_GW}

Following Groot and Warren,\cite{Gro:1997} we analyze the scaling behavior 
of the conservative interaction parameter $a$ by relating the
thermodynamic definition of the isothermal compressibility $\kappa_T$ to
the equation
of state (involving $a$) of a system of DPD-particles. The isothermal
compressibility is defined as the fractional change in 
the volume $V$
that results from a change in the pressure $P$ of the system, in a process
where the temperature $T$ is constant:
\begin{equation}
    \kappa_T = - \left.\frac{1}{V} \frac{\partial V}{\partial P}\right|_{T}
        = \left.\frac{1}{\rho} \frac{\partial \rho}{\partial P}\right|_{T}.
\end{equation}
Here $\rho$ denotes the number density of particles, i.e. $N/V$. It is 
convenient to define the dimensionless parameter $\kappa^{-1}$ by
\begin{equation}
    \kappa^{-1} = \frac{1}{k_B T \rho \kappa_T}
        = \frac{1}{k_B T} \frac{\partial P}{\partial \rho}.
\end{equation}
which we require to be invariant under scaling.
\begin{equation}
\label{requirement}
    \left. \frac{1}{k_B T \rho \kappa_T} \right|_\text{sim}
    = \left. \frac{1}{k_B T} \frac{\partial P}{\partial \rho}
        \right|_\text{sim}
    =  \left. \frac{1}{k_B T} \frac{\partial P}{\partial n}
        \right|_\text{exp} = \text{const.}
\end{equation}
Here, $n=\nu\rho$ denotes the molecular number density of the
physical system and the subscript ``$\text{exp}$'' relates to the 
experimental value, while ``$\text{sim}$'' refers to the simulation value.

The equation of state relates the pressure with the particle number density
$\rho$. From simulation results, Groot and Warren\cite{Gro:1997} evaluated
the virial expression
\begin{align}
    \label{eq_virial}
    P &= \rho k_B T + \frac{1}{3V} 
        \left< \sum_{j>i} r_{ij} F^C_{ij} \right>, \nonumber \\
      &= \rho k_B T +  \frac{2 \pi}{3} \rho^2 \int_0^{r_c}
        r a \left(1-\frac{r}{r_c}\right) g(r) r^2 dr.
\end{align}
Here, $g(r)$ denotes the radial distribution function. For densities
$\rho > 2$ in reduced units (i.e. more than 2 particles in a 
cubic box with
linear dimensions of $r_c$), the following equation of state is a 
good approximation to the numerical simulations:\cite{Gro:1997}
\begin{equation}
\label{EOS}
    P = \rho k_B T + a \alpha \rho^2, \qquad (\alpha = 0.101 \pm 0.001).
\end{equation}
From this, one concludes that the part of the pressure caused by the
conservative interaction scales linearly in $a$. From 
Eqn.~(\ref{requirement}), we obtain that at constant temperature
\begin{equation}
    \frac{1}{k_B T} \frac{\partial}{\partial \rho}
        \left( k_B T \rho + a \alpha \rho^2 \right)
    = \text{const.}
\end{equation}
Using $\rho=n/\nu$ where $n$ is the molecular number density and 
$\nu$ the chosen coarse-graining parameter, it follows that
\begin{equation}
\label{equ_a_relation}
    1 + \frac{2 a \alpha n}{k_B T \nu} = \text{const.}
\end{equation}
Since $n$ and $T$ are constant, Groot and Rabone\cite{Gro:2001} concluded 
that $a$ must scale linearly with $\phi=\nu'/\nu$ 
in order to maintain isothermal compressibility under
a change of the coarse-graining level.

\begin{figure}
    \centering
    \includegraphics[width=\columnwidth]{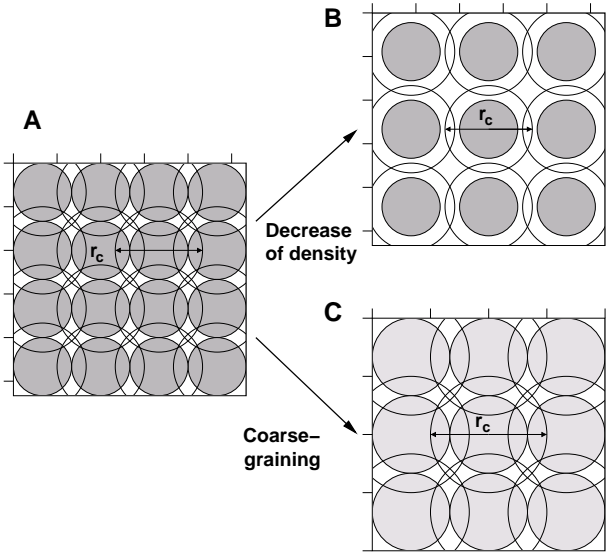}
    \caption{
Schematic of the scaling process: Frame~A shows a DPD simulation with a
cutoff radius of $r_c$. Frame~B depicts the coarse-graining procedure
performed in Groot and Rabone\cite{Gro:2001}. With changing particle
density, the particle diameter is kept constant while the
interparticle force is increased to maintain the system pressure. However,
in order to properly conserve systemic parameters like
compressibility, both the interaction parameter and the interaction cut-off
radius need to be increased as measured in physical units. Frame C~depicts
the proper scaling in coarse-graining. Along with a decrease of
the particle density (in physical units), the  interaction range is
increased. In this case, the interaction parameter $a$
scales differently than in Frame B in order to preserve systemic properties.
}
    \label{fig_coarsegraining}
\end{figure}

Several authors regard this scaling as an inherent drawback of the
DPD-method, since on the micrometer scale the method would appear to
be effectively thwarted. We claim that this scaling -- and hence its 
implications -- is wrong. The problem is rooted in the construction of 
the equation of state (Eqn.~\ref{EOS}). In their computer simulations, 
Groot and Rabone\cite{Gro:2001} decreased the
density of DPD particles while keeping the interaction cutoff radius $r_c$
constant. This approach allows one to keep the system behavior invariant by
scaling the interaction parameters shile changing the DPD particle 
density, without altering their properties. However, proper 
scaling means lowering
the number of employed DPD particles while simultaneously enlarging their
interaction radius. The difference is illustrated in 
Fig.~\ref{fig_coarsegraining}. Frame~\ref{fig_coarsegraining}A 
is taken to be
a system with fine coarse-graining. Frame~\ref{fig_coarsegraining}B 
represents a scaled system with a lower DPD particle density but 
unchanged 
particle diameters. The result is that the mutual overlap of the soft
particles is smaller (as seen in Frame~\ref{fig_coarsegraining}B).
Hence it is intuitively clear that the interaction parameter has to be 
increased in order to keep the system properties constant;
formally, this argument is reflected in Eqn.~(\ref{equ_a_relation}).
In contrast, Frame~\ref{fig_coarsegraining}C shows
the system with the same scaling ratio as for 
Frame~\ref{fig_coarsegraining}B, but with the {\em relative} overlap of 
the interacting particles kept constant, which is accomplished by
scaling $r_c$. A closer examination of 
Frame~\ref{fig_coarsegraining}C
shows that it is part of a magnified version of 
Frame~\ref{fig_coarsegraining}A, namely a system where all the lengths 
associated with a single DPD particle have been uniformly scaled by a 
factor $\phi$ while keeping the system size constant ($L'=L$). This results 
in the following scaling relations for the coarse-graining level, number, 
mass, and size of DPD particles:
\begin{align}
\label{eqn_scaling}
    \nu' &= \phi \nu \nonumber \\
        N'   &= \phi^{-1} N \nonumber \\
        m'   &= \phi\, m \nonumber\\
        r_c' &= \phi^{1/d}\, r_c,
\end{align}
where $d$ is the number of dimensions of the system (see 
Fig.~\ref{fig_coarsegraining}C).

\subsection{Scaling of the Potential Energy}
\label{sec_renorm_compress}

We start by calculating the change of potential energy $U$ of a system of
DPD particles enclosed in a box that undergoes compression. This change
is related to the compressibility of the system and is required to be
invariant under scaling. In practice, we require the dependence of 
$a$ on $\phi$, such that the chosen coarse-graining level does not affect 
the compressibility. For the uncompressed system, we have
\begin{equation}
    \label{eq_potenergy}
    U_0 = \sum_{i>j} \frac{\chi_{ij} a}{2 r_c} (r_{ij} - r_c)^2.
\end{equation}
This equation holds for soft core repulsions which are used 
throughout the literature. In general, the potential may be viewed as a 
harmonic approximation of any potential close to an energy minimum. For an
isotropically compressed system with box length  $(1-\epsilon)L$,
where $\epsilon \ll 1$ is the relative compression parameter, the
change in the interparticle distance $\Delta r_{ij}(\epsilon)$ is not 
assumed to be the same for all pairs of particles. However, we 
require that
\begin{equation}
    \Delta r_{ij}(\epsilon) = \epsilon r_{ij} + \mathcal{O}(\epsilon^2),
\end{equation}
which means that we rule out (first order) phase transitions
under compression. The total potential energy of the compressed system
is then given by
\begin{equation}
    U_\epsilon = \sum_{i>j} \frac{\chi_{ij} a}{2 r_c} (r_{ij}
    - \Delta r_{ij}(\epsilon) - r_c)^2.
\end{equation}

To first order in $\epsilon$, we obtain for the change of internal energy
\begin{equation}\label{eq_diffU}
    \Delta U = U_\epsilon - U_0
    = \sum_{i>j} \chi_{ij}\, a\, (1-\frac{r_{ij}}{r_c})\,\epsilon\,r_{ij}.
\end{equation}
Because the change in internal energy of the system as a whole has to be
invariant under scaling, we have
\begin{equation}\label{eq_diffUb}
    \sum_{i>j}^N \chi_{ij}a\left(1-\frac{r_{ij}}{r_c}\right)\epsilon r_{ij}
    = 
    \sum_{i>j}^{N'} \chi_{ij}a'
        \left(1-\frac{r'_{ij}}{r'_c}\right)\epsilon r'_{ij}
\end{equation}
Due to the scaling of $N$, the number of terms in the sum of the left hand 
side of Eqn.~(\ref{eq_diffUb}) is proportional to $\phi^{-1}$. Since we 
require $\Delta U$ to be invariant under scaling, the force constant $a$ 
has to scale as 
\begin{equation}
    \label{eqn_scaling_a}
    a' = \phi^{1-1/d} a,
\end{equation}
the $\phi^1$ coming from the change in the number of terms in the sum and 
the $\phi^{-1/d}$ from the change in length scale.
We note that this scaling differs from that found by Groot and Rabone
where $a$ scaled linearly with $\phi$ (see Eqn.~(\ref{requirement})).
This scaling is the result of our requirement of maintaining the fractional
particle overlap during the change of the coarse-graining level.
Everything else being equal, this scaling would seemingly still
imply an upper coarse-graining limit, although not as
severe as the initial result of Groot and Rabone\cite{Gro:2001}.
However, scaling affects not only length scales and the interaction
parameter, but also the energy and time scales implicit in the simulation.

\subsection{Scaling of the Kinetic Energy}
\label{sec_renorm_heat}

Since the unprimed and the primed
systems should be physically equivalent, we further require that $T'=T$ 
and $Q'=Q$, where $Q$ is the thermal energy, i.e.~heat content, of the 
whole system; with these requirements, physical properties become 
independent of the coarse-graining, and the coarse-graining level is 
solely a simulation parameter. In order to be consistent with statistical 
mechanics, a proper consideration of the reduction of the number of 
degrees of freedom resulting from coalescing several physical particles 
into a DPD particle is required. The heat content in a 3-D system 
consisting of particles with no internal structure is given by
\begin{equation}
	\label{eqn_heat}
    Q_{\text{phys}} = \frac{f_\text{phys}}{2} k_B T  
        = \frac{N_\text{phys}}{2}
            \expt{m_\text{phys} \mathbf{v}_\text{phys}^2},
\end{equation}
where  $f_\text{phys} = d N_\text{phys}$ is the number of degrees of
freedom. Forming $N$ DPD-particles, i.e.~coherently moving groups of $\nu$
physical particles, changes the relation between kinetic
energy, temperature and number of particles. It holds
\begin{equation}\label{eqn_kdef}
   Q = \frac{f}{2} k T = \frac{N}{2}\expt{m \mathbf{v}^2}
\end{equation}
with $f = 3N, k = \nu k_B, N = N_\text{phys} / \nu, m = \nu m_\text{phys}$. 
Consequently, from the required invariance $Q'=Q=Q_{\text{phys}}$ 
and $T'=T$ for two simulations, we get a scaling relation
\begin{equation}
    \label{eqn_boltzmann}
    k' = \phi k
\end{equation} 
We introduce the parameter $k$ that has the role of a Boltzmann constant
for systems with reduced number of degrees of freedom. (This scaling 
affects the dissipation-fluctuation relation 
Eqn.~(\ref{eq_dissipation_fluctuation2}) which now reads
$\sigma^2=2kT\gamma$)
That Eqn.~(\ref{eqn_kdef}) and hence  $Q=Q'$ is consistent with the DPD 
method is discussed and explicitly corroborated in 
Sec.~\ref{sec_simulation}.

The behavior of $\gamma$ and $\sigma$ under scaling remains to be 
determined. Examining Eqn.~(\ref{eq_virial}), we note that the pressure 
is independent of $\gamma$ and $\sigma$. This means that, with respect to 
static compressibility, we have significant freedom in the choice of the 
scaling function $\Gamma$:
\begin{align}
\label{eqn_gammascaling}
	\gamma'   &= \Gamma(\phi) \gamma \nonumber \\
	\sigma'   &= (\Gamma(\phi) \phi)^{1/2} \sigma
\end{align}
the latter equation is a consequence of $\sigma' = \sqrt{2 \gamma' k' T'}$. 
Dimensional analysis motivates the choice
\begin{equation}
    \Gamma(\phi) = \phi^{1-1/d},
\end{equation}
which in turn implies that
\begin{eqnarray}
    \label{eqn_gamma_sigma_scaling}
    \gamma' &=& \phi^{1-1/d}\gamma \nonumber \\
    \sigma' &=& \phi^{1-2/d}\sigma.
\end{eqnarray}
This specific choice will later be shown to be crucial for establishing 
the scalability of the method, but investigations with other 
goals (such that deciding on a coarse-graining level $\nu$ with least
artefacts) may require alternative gauges.

To summarize what has been established throughout the last two sections,
changing the level of coarse-graining in DPD requires the following
scaling relations:
\begin{equation}
\label{eqn_scaling_summary}
    \begin{array}{rclcrcl}
	N'   &=& \phi^{-1} N       & \quad & a'      &=& \phi^{1-1/d} a \\
	m'   &=& \phi\, m          & \quad & \gamma' &=& \phi^{1-1/d}\gamma \\
	r_c' &=& \phi^{1/d}\, r_c  & \quad & \sigma' &=& \phi^{1-2/d}\sigma \\
    k'   &=& \phi k
    \end{array}
\end{equation}

\subsection{Experimental validation}
\label{sec_simulation}

In order to illustrate the correctness of the above scaling arguments, we
have measured the pressure in simulations as a function of $a$ for
different coarse-graining levels and temperatures. In these simulations,
the DPD particles are confined to a box with hard walls. When a particle 
collides with a wall, it is reflected elastically and the instantaneous
impulse normal to the wall is measured. The pressure is measured as the 
time-average of the normal forces on the walls divided by the 
surface area of the cube: 
$P = \langle m\Delta \mathbf v_\bot / (A\Delta t) \rangle$ where
$A$ is the area of the box and $\Delta \mathbf v_\bot$ is the component of 
the particle velocity orthogonal to the wall.

\begin{figure}
\centering
\psfrag{xlabel}[t][]{$a/\phi^{1-1/d}$}
\psfrag{ylabel}[][]{Pressure $P$}
\includegraphics[width=\columnwidth]{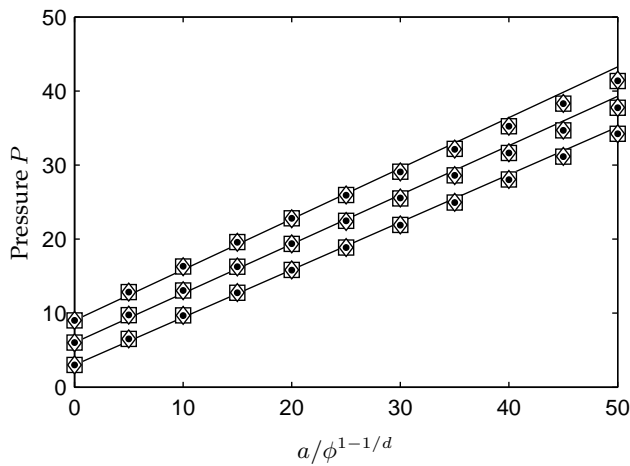}
\caption{
    \label{fig:pressure}
    The calculated pressure as a function of the interaction strength $a$.
    We show the pressure for the basic DPD model, corresponding to
    $r_\text{c}=1$ (white squares), and a rescaled simulation of the
    same 3D system with $\phi=8$ corresponding to $r_c'=2r_c$ (diamonds
    with black dots). Particle density and box length are set to 
    $3r_c^{-3}$, $10r_c$, and ${3r_c'}^{-3}$, $10r_c'$, respectively.
    Different temperatures $T_1$, $T_2$, and $T_3$ are considered, such 
    that $kT_1=1$, $kT_2=2$, $kT_3=3$, and $k'T_1'=1$, $k'T_2'=2$,
    $k'T_3'=3$, respectively
    (bottom to top). Solid lines show the prediction of the equation of 
    state (Eqn.~(\ref{EOS})) with $\alpha$ estimated from the simulation 
    results (for the case of $r_c=1$). The pressure increases approximately
    linearly for $a$ not too large. Furthermore, the values for the
    original and the rescaled simulations coincide.
}
\end{figure}

In Fig.~\ref{fig:pressure} we show a comparison between $r_c = 1$ (white
squares), corresponding to normal DPD, and a
rescaled system with $r'_c = 2r_c$, corresponding to $\phi = 8$
diamonds with black dots). Particle density and box length are set to 
$3 r_c^{-3}$, $10r_c$, and $3r_c'^{-3}$, $10r_c'$, respectively. 
Different temperatures $T_1$, $T_2$, and $T_3$ are considered, such that
$kT_1=1$, $kT_2=2$, $kT_3=3$, and $k'T_1'=1$, $k'T_2'=2$, $k'T_3'=3$, 
respectively
(bottom to top in Fig.~\ref{fig:pressure});
this is a slight deviation from the conventional usage of reduced units.
For the rescaled simulation, all parameters ($N$, $m$, $r_c$, $a$,
$k$, $\gamma$, and $\sigma$) have been scaled according to
Eqn.~(\ref{eqn_scaling_summary}). We want to show that the pressure
of this system is invariant under scaling, for all values of $a$
and $T$, if we follow the described scaling relations.

In Fig.~\ref{fig:pressure}
the pressure is plotted as a function of $a/\phi^{1-1/d}$, so that we
should obtain the same curve independent of the coarse-graining level.
On the one hand, for $a=0$, the case of an ideal gas, the pressure
should be given by
\begin{equation}
    \left.P\right|_{a=0} = \phi^{-1}N \phi k T / V = \rho kT,
    \label{eqn_idealgas}
\end{equation}
which it indeed is seen to be. On the other hand, kinetic gas theory 
establishes 
\begin{equation}
    P=\rho\langle m \mathbf v^2\rangle/3.
    \label{eqn_kineticgas}
\end{equation}
This constitutes a relationship between the thermostat and the
conservative mechanical interactions with the walls, which is non-trivial
for our simulation. 
The way we measure the pressure gives a direct relation to 
$\langle\mathbf v^2\rangle$. 
The pressure calculated from kinetic quantities 
(Eqn.~(\ref{eqn_kineticgas})) coincides with the value required from the 
thermodynamic relation Eqn.~(\ref{eqn_kdef}). This shows that 
Eqn.~(\ref{eqn_kdef}) holds for our simulation and thereby confirms the 
scaling relation for $k$ given in Eqn.~(\ref{eqn_boltzmann}).

Eqn.~(\ref{EOS}) estimates a linear relation between $a$ and $P$.
This relation is shown as solid lines, whereby the Boltzmann factor
in Eqn.~(\ref{EOS}) has been replaced by its scaled version $k$. For each 
line, $\left.P\right|_{a=0}$ is the theoretical value and the slope is
the average obtained from each data set $\left(P(a)-P(0)\right)/a$.
Unless $a$ is too large, the pressure increases linearly with $a$.
For high values of $a$, one finds that Eqn.~(\ref{EOS}) is not a good
approximation to the measured pressure, anymore.

The plot also shows that the scaling of $a$ is correct as the pressure
values of the original and the rescaled systems fall right on top of
each other, regardless of the value of $a$. The conclusion from these
comparisons is that the pressure is indeed invariant when we use the 
scaling relations given in Eqn.~(\ref{eqn_scaling_summary}).

\section{Reduction of Units}
\label{sec_reduced_units}

Our goal is now to show that the velocity increments  $\Delta \mathbf v$
obtained from integrating the forces are unchanged when the scaling is
combined with the according reduction of units: 
$\Delta\mathbf{\tilde{v}} = \Delta\mathbf{v'}$, which
implies that the relative particle motions are unaffected by
scaling in the reduced unit systems.

In the DPD literature, length, mass, and energy are considered primary
units, leading to a derived unit of time $\tau$ given by
\begin{equation}
	\tau = r_c \sqrt{\frac{m}{k T}}\:,
\end{equation}
From the above arguments we obtain the time-scale in the rescaled system:
\begin{equation}
	\label{eqn_time_scale}
	\tau' = r_c' \sqrt{\frac{m'}{k' T'}}
	= \phi^{1/d} r_c \sqrt{\frac{\phi m}{\phi k T}}
	= \phi^{1/d}\, \tau .
\end{equation}
Thus, length and time scale in the same way. This also implies
that the velocities are invariant as they are given by $\Delta r/\Delta t$.

The random variable $\zeta_{ij}$ has the unit $\tau^{-1/2}$, as
noted in the discussion following Eqn.~\ref{equ_basicDPD}.
Given the scaling of $\tau$, it follows that
\begin{equation}
    \zeta_{ij}' = \phi^{-1/(2d)} \zeta_{ij},
\end{equation}
and therefore
\begin{equation}
    \sigma'\zeta_{ij}' = \phi^{1-1/d} \sigma\zeta_{ij}.
\end{equation}

Since the other terms in Eqn.~(\ref{equ_basicDPD}) are all scale-free,
the three force components of DPD all scale by a factor $\phi^{1-1/d}$.

When velocity increments are calculated during one time step, one finds
that the force scaling is canceled by the scaling of mass and time. This
is shown below for the conservative force:
\begin{align}
\label{eq_v_c_invariant}
    [\Delta \mathbf v_i^C]'
        &= \sum_{j \ne i} \frac{\Delta t' \left[\mathbf F_{ij}^C\right]'}
            {m'}
         = \frac{a'}{m'} \, \Delta t' \, \sum_{j \ne i} \chi_{ij}
            \left(1 - \frac{r'_{ij}}{r_c'} \right) \mathbf{\hat r}_{ij}
            \nonumber\\
        &= \frac{\phi^{1-1/d}\phi^{1/d}}{\phi} \frac{a}{m}
            \, \Delta t  \, \sum_{j \ne i} \chi_{ij}
            \left(1 - \frac{\phi^{1/d}r_{ij}}{\phi^{1/d}r_c} \right)
                \mathbf{\hat r}_{ij}
            \nonumber\\
        &= \Delta \mathbf v_i^C .
\end{align}
The calculations for $[\Delta \mathbf v_i^D]'$ and
$[\Delta \mathbf v_i^R]'$ give similar results:
\begin{align}
\label{eq_v_d_invariant}
    [\Delta \mathbf v_i^D]'
        &= -\sum_{j \ne i} \frac{\gamma'}{m'} \, \omega^D(r_{ij}')
            [(\mathbf v_i'-\mathbf v_j')\cdot \mathbf{\hat r}_{ij}]
            \mathbf{\hat r}_{ij} \, \Delta t' \nonumber \\
        &= -\sum_{j \ne i} \frac{\phi^{1-1/d }\phi^{1/d}}{\phi}
            \frac{\gamma}{m} \, \omega^D(r_{ij})
            [(\mathbf v_i-\mathbf v_j)\cdot \mathbf{\hat r}_{ij}]
            \mathbf{\hat r}_{ij} \, \Delta t \nonumber \\
        &= \Delta \mathbf v_i^D \\
\label{eq_v_r_invariant}
    [\Delta \mathbf v_i^R]'
        &= \sum_{j \ne i} \frac{\sigma'}{m'}
            \omega^R(r_{ij}') \zeta_{ij}'\mathbf{\hat r}_{ij}
            \, \Delta t' \nonumber \\
        &= \sum_{j \ne i} \frac{\phi^{1-1/(2d)}\phi^{-1/(2d)}\phi^{1/d}}
            {\phi}\frac{\sigma}{m} \omega^R(r_{ij}) \zeta_{ij}
            \mathbf{\hat r}_{ij}\, \Delta t \nonumber \\
        &= \Delta \mathbf v_i^R
\end{align}
Since $\Delta \tilde{\mathbf r} = \Delta \mathbf r' / r'_c$ and
$\Delta \tilde t = \Delta t' / \tau'$, we get for the velocity increment
by considering
\begin{equation}
    \label{eqn_v_tilde}
    \Delta \tilde{\mathbf v}_i = \Delta \tilde{\mathbf r}_i / \Delta
    \tilde t
    = \Delta \mathbf v'_i \: \tau' / r_c' .
\end{equation}
Because time and length scale in the same way we get
$\tau'/r_c' = \tau/r_c$. Combining this with
$\Delta \mathbf{v} = \Delta \mathbf{v}'$
(Eqns.~(\ref{eq_v_c_invariant}) to (\ref{eq_v_r_invariant})) one finally
obtains
\begin{equation}
    \label{eqn_v_reduced}
    \Delta \tilde{\mathbf v}_i = \Delta \mathbf v_i,
\end{equation}
which implies
\begin{equation}
    \mathbf{\tilde r}(\tilde t) = \mathbf{r}(t).
\end{equation}

What remains to be shown is the scaling of the reduced parameters
$\tilde a$, $\tilde \gamma$, and $\tilde \sigma$. Since $a$  scales like
energy over length, when going to the reduced units of the primed system,
we have
\begin{equation}
    \tilde a = a' \frac{r_c'}{k'T'}
        = \frac{\phi^{1-1/d}\phi^{1/d}}{\phi} a \frac{r_c}{k T}
        = a ,
\end{equation}
and similarly, since $\gamma$ scales like energy over length and velocity,
from $\tilde \gamma = \gamma' \, r_c'^2/(k'T'\tau')$ we get
\begin{equation}
    \tilde \gamma
     = \gamma' \frac{r_c'^2}{k'T'\tau'}
        = \frac{\phi^{1-1/d}\phi^{2/d}}{\phi\phi^{1/d}}
            \gamma \frac{r_c^2}{kT\tau}
        = \gamma .
\end{equation}
From the fluctuation-dissipation relation it follows again that
\begin{equation}
    \tilde \sigma = \sigma .
\end{equation}
Hence, scaling and unit reduction precisely cancel each other. As a result,
the DPD formalism is scale-free. This means that the calculation 
with one and same set of parameter values represents systems at arbitrary 
lengths scales.

Note that in order to understand the according physical time scales we have 
to comment on transport properties of the method and the scaling
of fluctuations. We base our argument on diffusion, but could equally well
consider viscosity since the two are related by the Schmidt-number, which
is dimensionless. The numerical equivalence of the measured diffusion
constants $\tilde D = D$ (being a consequence of the equality of the
measured displacement) and the fact that diffusion scales like length
squared over time causes an apparent problem: it seemingly implies that
relative fluctuations $\tilde{D}/\tilde{L}$ stay constant 
instead of vanishing. This problem disappears when one calibrates the 
simulation to an actual physical system. Assume that the cutoff radius is 
related to a physical length by $r_c = l [\text cm]$. 
We then have
\begin{equation}
D \frac{r_c^2}{\tau} = D_{\text{phys}} \frac{\text{cm}^2}{\text{sec}},
\end{equation}
with $D_{\text{phys}}=\frac{\sqrt{<r(t)^2>}} t$
referring to the diffusion constant in physical units. We get
\begin{equation}
\tau = \frac{D}{D_{\text{phys}}} l^2 [\text{sec}].
\end{equation}
Consequently, expressed in physical units, it holds for the fluctuations
\begin{equation}
\lim_{l \to \infty} \frac{D_{\text{phys}}}{L_{\text{phys}}} \sim
    \frac 1 l = 0.
\end{equation}

\section{Summary and Discussion}

We have shown that each individual term of a reduced unit DPD calculation 
is scale-free, and hence so is DPD as a whole. We emphasize that this 
result was only achieved by a combined effect of scaling, the reduction 
of units, and finally by requiring a specific but obvious scaling of 
$\gamma$ and $\sigma$. 

The scaling of energy per particle (or equivalently the number of degrees
of freedom in the system) is shown to
require increasing the value of $k$ by the same factor of $\phi$
that the number of particles is reduced by. This scaling of $k$
is required to keep the total system energy constant. The increase in $k$
compensates for the freezing effect seen by other authors who failed to
take this into account and who also did not uniformly scale all length
parameters.

In this article, we have shown that the DPD-method is scale-free for the 
simulation of bulk fluids. This is not necessarily the case e.g. for binary
mixtures of liquids $A$ and $B$ where several conservative interaction
parameters occur, e.g. $a_{AA},a_{AB},a_{BB}$. Whereas bulk interactions
given by $a_{AA},a_{BB}$ scale as discussed in
this article, $a_{AB}$ is a surface term that determines interfacial energy
and therefore scales differently. This is, however, beyond the scope of
this article.

\begin{acknowledgments}
This work was financially supported by the EU project FP6 IST-FET integrated
project PACE and by the Los Alamos National Laboratory LDRD-DR grant on
``Protocell Assembly'' (PAs). The first author was additionally supported 
by the BMBF project Systems Biology of the Liver Grant \# 3P3137. Most of 
the work was performed during a workshop at the European Center for Living 
Technologies (ECLT) in Venice, Italy, organized by Martin Nilsson Jacobi. 
The stimulating working environment provided by this institution is greatly 
acknowledged.
\end{acknowledgments}

\bibliography{references}

\end{document}